%% file: main.tex
\documentclass[runningheads]{llncs}
\usepackage{graphicx}
\usepackage{amssymb}
\usepackage{xcolor}
\usepackage{subfig}
\usepackage[utf8]{inputenc}
\usepackage{fdsymbol}
\usepackage{graphicx}

\newcommand{\ttt}{\boldsymbol{\theta}}

\begin{document}
\title{Challenging mitosis detection algorithms: \\ Global labels allow centroid localization\thanks{The work of C. Fernández-Martín and U. Kiraz was funded from the Horizon 2020 of European Union research and innovation programme under the Marie Sklodowska Curie grant agreement No 860627 (CLARIFY Project). The work of Sandra Morales has been co-funded by the Universitat Politècnica de València through the program PAID-10-20. This work was partially funded by GVA through project PROMETEO/2019/109. }}
\titlerunning{Challenging mitosis detection algorithms}

\author{
Claudio Fernandez-Martín\inst{1} \and
Umay Kiraz\inst{2, 3} \and
Julio Silva-Rodríguez\inst{4} \and
Sandra Morales\inst{1} \and
Emiel Janssen\inst{2, 3} \and
Valery Naranjo\inst{1}
}

\authorrunning{C. Fernandez-Martín et al.}

\institute{
Institute of Research and Innovation in Bioengineering, \textit{Universitat Polit\`ecnica de Val\`encia}, Valencia, Spain (\email{clferma1@i3b.upv.es})\\
\and
Department of Pathology, \textit{Stavanger University Hospital}, Stavanger, Norway\\
\and
Department of Chemistry, Bioscience and Environmental Engineering, \textit{University of Stavanger}, Stavanger, Norway
\and
Institute of Transport and Territory, \textit{Universitat Polit\`ecnica de Val\`encia}, Valencia, Spain\\
}

\maketitle              % typeset the header of the contribution
\begin{abstract}

Mitotic activity is a crucial proliferation biomarker for the diagnosis and prognosis of different types of cancers. Nevertheless, mitosis counting is a cumbersome process for pathologists, prone to low reproducibility, due to the large size of augmented biopsy slides, the low density of mitotic cells, and pattern heterogeneity. To improve reproducibility, deep learning methods have been proposed in the last years using convolutional neural networks. However, these methods have been hindered by the process of data labelling, which usually solely consist of the mitosis centroids. Therefore, current literature proposes complex algorithms with multiple stages to refine the labels at pixel level, and to reduce the number of false positives. In this work, we propose to avoid complex scenarios, and we perform the localization task in a weakly supervised manner, using only image-level labels on patches. The results obtained on the publicly available TUPAC16 dataset are competitive with state-of-the-art methods, using only one training phase. Our method achieves an F1-score of $0.729$ and challenges the efficiency of previous methods, which required multiple stages and strong mitosis location information.

\end{abstract}

\input{sections/1_introduction}

\input{sections/2_related_work}
\input{sections/3_methods}
\input{sections/4_experiments}
\input{sections/5_results}
\input{sections/6_conclusions}


\bibliographystyle{IEEEbib}
\bibliography{references}

%Author, F.: Article title. Journal \textbf{2}(5), 99--110 (2016)
%Author, F., Author, S.: Title of a proceedings paper. In: Editor,
%F., Editor, S. (eds.) CONFERENCE 2016, LNCS, vol. 9999, pp. 1--13.
%Springer, Heidelberg (2016). \doi{10.10007/1234567890}
%Author, F., Author, S., Author, T.: Book title. 2nd edn. Publisher,
%Author, A.-B.: Contribution title. In: 9th International Proceedings
%on Proceedings, pp. 1--2. Publisher, Location (2010)
%LNCS Homepage, \url{http://www.springer.com/lncs}. Last accessed 4
%\end{thebibliography}

\end{document}

%% file: sections/1_introduction.tex
\section{Introduction}

\begin{figure}[h!]
    \begin{center}
    
            %\vspace{-1 em}

          \subfloat[\label{fig:dataset1a} Mitotic cells]{\includegraphics[width=0.6\linewidth]{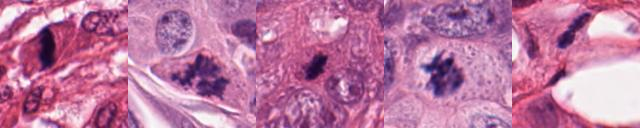}}
          \thinspace
          
          \vspace{-1 em}

          \subfloat[\label{fig:dataset1c} Non-mitotic cells]{\includegraphics[width=0.6\linewidth]{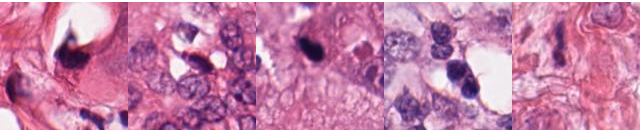}}
    
        \caption{Visual illustration of the morphological heterogeneity and the challenge of differentiating patterns between mitotic and non-mitotic cells, extracted from TUPAC16 \cite{tupac16}.}
    
        \label{fig:dataset}
    \end{center}
\end{figure}

In digital pathology, mitosis counting is one of the most important tasks in the histopathological clinical practice. In the case of breast cancer, the mitotic activity index (MAI) is considered one of the strongest proliferation-associated prognostic factors \cite{baak_2005}. However, mitosis counting is a laborious and time-consuming task due to the large size of the Hematoxylin and Eosin (H\&E) slides under a microscope, and the low occurrence of mitotic figures. In addition, the large heterogeneity of patterns and similarity between mitotic and non-mitotic cells (see Figure \ref{fig:dataset}) makes this task highly variable among clinical experts \cite{elmore_diagnostic}, which hinders its reproducibility.

\vspace{-2 em}

In the last years, the advent of modern deep learning algorithms has emerged as a possible solution to bring objectivity and reproducibility to the challenge of mitosis localization. Deep learning using convolutional neural networks (CNNs) has reached remarkable results in a wide range of applications under the supervised learning paradigm. Nevertheless, it requires a reasonable amount of carefully-labeled data to perform properly. In the case of mitosis localization, this is a tedious process, which usually is repeated by different pathologists to reach consensus labels. Since delineating at pixel-level individual mitotic cells is an unfeasible task, the reference datasets normally contain centroid-based labels \cite{tupac16} or inexact pixel-level annotations \cite{mitos12}. Because of this, previous works to automate the mitosis localization process have struggled to match the available labels to the use of segmentation or object detection CNNs, which are typically used localization tasks. Contrary to this line of work, we propose to make use of inherent spatial localization capacity of CNNs in image-level classification tasks \cite{Oquab2014}, without the need to resort to an exact localization of the mitotic cell inside the region of interest. Our main contributions are summarized as follows:
\vspace{-0.5 em}
\begin{itemize}
\item A CNN for weakly supervised segmentation of mitotic figures on H\&E patches using image-level labels.
\item Concretely, training is driven by maximum aggregation of instance-level predictions.
\item Comprehensive experiments demonstrate the competitive performance on the popular TUPAC16 dataset, using a single-phase pipeline without requiring the exact localization information for training our model.
\end{itemize}

%% file: sections/2_related_work.tex
\section{Related Work}

\subsection{Mitosis detection}

Mitosis localization algorithms using CNNs deal with labels in the form of centroid annotations, or inexact pixel-level delineation of the mitotic cell. In that sense, Li \textit{et al.} \cite{li_concentric} propose a novel concentric loss to move from centroid labels to pixel-level segmentation using the pixels surrounding certain radius of the centroid. Other works focus on leveraging cell-level predictions using multi-phase pipelines \cite{paeng_2016,zerhouni_wide,akram_2018,wahab_2019,nateghi_2021,sohail_multiphase}. For instance, Sohail \textit{et al.} \cite{sohail_multiphase} propose a complex multi-phase pipeline that includes pseudolabeling centroid-labelled mitosis via previously trained Mask R-CNNs. Also, Nateghi \textit{et al.} use multiple training stages to refine the false positive detection using hard-negative mining via stain priors, or prediction uncertainty. In contrast to these works, we study how training a CNN at the image-level for a classification task also allows the precise location of mitotic cells without using any localization information, shape or stain priors, or multi-phase refinement pipelines.

\subsection{Weakly supervised segmentation}

Weakly supervised segmentation (WSS) aims to leverage pixel-level localization using global (a.k.a image-level) labels during training. According to \cite{Ilse2018}, WSS methods use fully-convolutional CNNs with an aggregation function that merges all the spatial information into one value, that serves as global prediction \cite{weglenet}. This output is then used to compute the loss function, and drives the network optimization. Different strategies include the aggregation of spatial features (embedding-based) or pixel-level predictions (instance-based). Finally, the probability maps before aggregation operation are used as segmentation predictions. Lately, these segmentation maps are refined to incorporate self-supervised learning pipelines \cite{Wang2020} or uncertainty proxies \cite{belharbi_wss}, among others.

%% file: sections/3_methods.tex
\section{Methods}

\begin{figure}[h!]
    \begin{center}
    \includegraphics[width=1\linewidth]{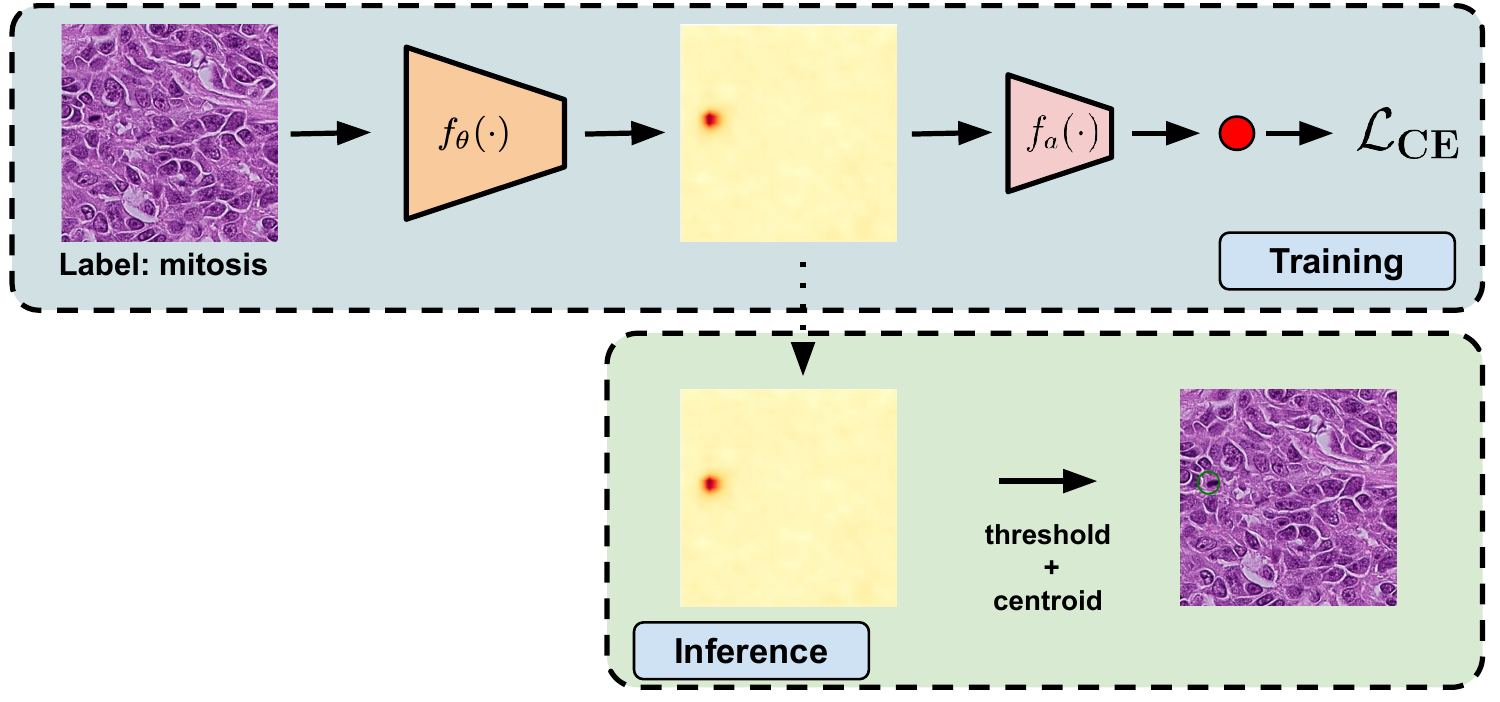}
    \end{center}
    \vspace{-2 em}
    \caption{Overview of the proposed method for mitosis localization.}
    \label{fig:summary}
    \vspace{-1 em}
\end{figure}

An overview of our proposed method is depicted in Figure \ref{fig:summary}. In the following, we describe the problem formulation, and each of the proposed components.

\paragraph{\textbf{Problem Formulation}} In the paradigm of weakly supervised segmentation (WSS), the training set is composed of images $\{x_n\}_{n=1}^{N}$, whose binary label $\{Y_{n}^k\}_{n=1}^{N}$, such that, $Y_{n}^k = \{0, 1\}$ is known, and defines if a category $k$ is present within the image. Also, each positive image has pixel-level labels $y_{n,i}$ for each $i$ pixel in the image, but they remain unknown during training. Further, we denote $Y_{n}^k$ as $Y_{n}$ for simplicity, since one unique class is taken into account, and we assume image index $n$.

\paragraph{\textbf{Instance-based WSS}}

In this work, we aim to train a CNN capable of locating positive mitosis during inference, while being trained only with image-level labels. To do so, we make use of an instance-based weakly supervised learning strategy. Let us denote a CNN model, $f_{\ttt}(\cdot) : \mathcal{X} \rightarrow \mathcal{H}^{K}$, parameterized by $\ttt$, which processes instances $x \in \mathcal{X}$ to output sigmoid-activated instance-level probabilities, $h_i$, such that $h_i \in [0, 1]$. Also, we use a parameter free aggregation function, $f_{a}(\cdot)$, in charge of combining the pixel-level scores into one global output $H$, such that $H = f_{a}(f_{\ttt}({x}))$. Then, the optimization of $\ttt$ is driven by the minimization of cross entropy loss between reference and predicted image-level score.

\vspace{-1 em}
\begin{equation}
\label{eq:ce}
\mathcal L_{ce} = Ylog(H) + (1-Y)log(1-H)
\end{equation}

In this work, we propose to use the maximum operation as aggregation function, $f_{a}(\cdot)$. Although this aggregation only backpropagates gradients through the maximum-activated spatial regions, this effect produces that only very discriminative cells will be classified as mitosis, which avoids false positive predictions.

\paragraph{\textbf{Inference}} During inference, pixel-level predictions are inferred using the pixel-level predictions given by the trained CNN, $y_{i} = f_{\ttt}(x)$. The probability maps are resized to the original image dimensions by bi-linear interpolation. Then, sigmoid scores are converted to a binary mask by applying a threshold to the probability maps. Concretely, the threshold is obtained from the operative point of the ROC curve between image-level predictions and references. Finally, a centroid is assigned to each element in the mask, to be located as a mitosis.

%% file: sections/4_experiments.tex
\section{Experimental setting}

\subsection{Datasets}

The experiments described in this work are carried out using the popular 2016 TUmor Proliferation Assessment Challenge (TUPAC16) dataset \cite{tupac16}. TUPAC16 is publically available and is composed of $73$ breast cancer whole slide images from two different institutions. In particular, the auxiliary mitosis dataset contains $1552$ processed regions of interest at $40\times$ magnification, with centroid-labelled mitosis by consensus of expert pathologists. Following relevant literature in \cite{li_concentric}, we extracted patches of size $500$ pixels from the regions of interest for computational efficiency. The dataset is divided into patient-level training, validation, and testing cohorts in a similar fashion to prior literature \cite{li_concentric}.

\subsection{Metrics}

We use standard metrics for mitosis localization evaluation. First, the model is optimized using only global image-level labels, by means of the accuracy, AUC, and F1-score. Then, the comparison with state-of-the-art methods on mitosis detection is assessed using the standard criteria of mitosis detection contests \cite{sohail_multiphase}. A detected mitosis is considered true if it is located at most 30 pixels from an annotated mitosis. Under this criteria, precision, recall and F1-score are computed.

\subsection{Implementation details}

The proposed method is trained using ResNet-18 \cite{He2016} convolutional blocks as a backbone. Concretely, the first $3$ blocks pre-trained on ImageNet are used as feature extractor, which are retrained for the mitosis detection task. We trained this architecture during $40$ epochs to optimize Eq. \ref{eq:ce} using a batch size of $32$ images and a learning rate of $0.0001$. In order to deal with class imbalance, the images are sampled homogeneously according to its class in each epoch. Also, color normalization and augmentation techniques are employed to increase robustness against stain variations and artifacts in the digitized slides. Images are color-normalized using the stain normalization method of Macenko \textit{et al.} \cite{macenko_stainnorm}, and data augmentation is included during training using spatial translations, rotations, and blurring.

%% file: sections/5_results.tex
\section{Results}

\subsection{Comparison to literature}

The quantitative results obtained by the proposed method for mitosis localization on the test cohort are presented in Table \ref{soa_comparison}. Also, we include results reported in previous literature on the TUPAC16 dataset. The proposed weakly-supervised method reaches an F1-score value of 0.729, which is comparable to prior literature without accessing to any supervision regarding the exact location of the mitosis in the image. It should be noted that, in addition, the best previous methods use additional training data, and require multiple stages of label refinement. In contrast, the proposed method uses only one training cycle. Moreover, the proposed approach obtains the best precision on mitosis localization that only use one training phase. This could be due to maximum aggregation, which propagates gradients only in those regions that are highly discriminating. 

\begin{table}[h!]
\resizebox{\linewidth}{!}{
\begin{tabular}{|l|c|c|c|c|c|c|}
\hline
\multicolumn{1}{|c|}{\textbf{Method}} &
  \textbf{Precision} &
  \textbf{Recall} &
  \textbf{F-score} &
  \multicolumn{1}{c|}{\textbf{\begin{tabular}[c]{@{}c@{}}Multiple\\ phases\end{tabular}}} &
  \textbf{\begin{tabular}[c]{@{}c@{}}Location \\ supervision\end{tabular}} &
  \multicolumn{1}{c|}{\textbf{\begin{tabular}[c]{@{}c@{}}External\\ data\end{tabular}}} \\ \hline
Paeng \textit{et al.} (2017) \cite{paeng_2016}    & -     & -     & 0.652 & $\color{black}\times$ & $\color{black}\times$ &  \\ \hline
Zerhouni \textit{et al.} (2017) \cite{zerhouni_wide} & 0.675 & 0.623 & 0.648 & $\color{black}\times$ & $\color{black}\times$ &  \\ \hline
Akram \textit{et al.} (2018) \cite{akram_2018}    & 0.613 & 0.671 & 0.640 & $\color{black}\times$ & $\color{black}\times$ & $\color{black}\times$ \\ \hline
Li \textit{et al.} (2019) \cite{li_concentric}       & 0.64  & 0.70   & 0.669 &  & $\color{black}\times$ &  \\ \hline
Wahab \textit{et al.} (2019) \cite{wahab_2019}   & 0.770 & 0.660 & 0.713 & $\color{black}\times$  & $\color{black}\times$ &  \\ \hline
Mahmood \textit{et al.} (2020) \cite{mahmood_2020}  & 0.641 & 0.642 & 0.642 & $\color{black}\times$ &  &  \\ \hline
Nateghi \textit{et al.} (2021)  \cite{nateghi_2021}  & 0.764 & 0.714 & 0.738 & $\color{black}\times$ & $\color{black}\times$ &  \\ \hline
Sohail \textit{et al.} (2021) \cite{sohail_multiphase}  & 0.710 & 0.760 & 0.750 & $\color{black}\times$ & $\color{black}\times$ & $\color{black}\times$ \\ \hline
Proposed               & 0.739 & 0.720 & 0.729 & &  &  \\ \hline
\end{tabular}
}
\caption{Performance comparison of the proposed model with existing methods on test subset of TUPAC16 auxiliary dataset.}
\label{soa_comparison}
\vspace{-2 em}
\end{table}

\vspace{-2 em}
\subsection{Ablation experiments}

In the following, we depict ablation experiments to motivate the choice of the different components of the proposed method.

\paragraph{\textbf{Weakly Supervised Setting}.}First, we study the configuration of the WSS model architecture. To do so, we explore the most outstanding configurations. First, embedding-based approaches that aggregate spatial features before the classification layers, and instance-based approaches that apply the classification layer spatially. Also, we use different aggregation methods, such as mean and max operations, and the trainable attentionMIL mechanism  \cite{Ilse2018}. Results are presented in Table \ref{ablation_wsss}. The figures of merit show that, although all methods reach similar results at image-level, only the instance-based with maximum aggregation performs properly on mitosis localization, since it is the only method that penalizes false positive localization during training.

\begin{table}[htb]
\centering
\begin{tabular}{|l|c|c|}
\hline
\multicolumn{1}{|c|}{\textbf{Configuration}} & \multicolumn{1}{c|}{\textbf{\begin{tabular}[c]{@{}c@{}}F-score\\ image-level\end{tabular}}} & \multicolumn{1}{c|}{\textbf{\begin{tabular}[c]{@{}c@{}}F-score\\ localization\end{tabular}}} \\ \hline
embedding - mean & 0.762 & 0.134 \\ \hline
embedding - max  & \textbf{0.772} & 0.234 \\ \hline
attentionMIL \cite{Ilse2018}    & 0.768  &  0.014 \\ \hline
instance - mean  & 0.753 & 0.004 \\ \hline
instance - max   & 0.761 & \textbf{0.729} \\ \hline
\end{tabular}
\caption{Performance comparison of the different configurations of the WSS proposed model, in terms of aggregation strategies. Results are presented for mitosis localization and image-level classification.}
\label{ablation_wsss}
\end{table}
\vspace{-2 em}

\begin{figure}[htb]
    \begin{center}
    \includegraphics[width=.7\linewidth]{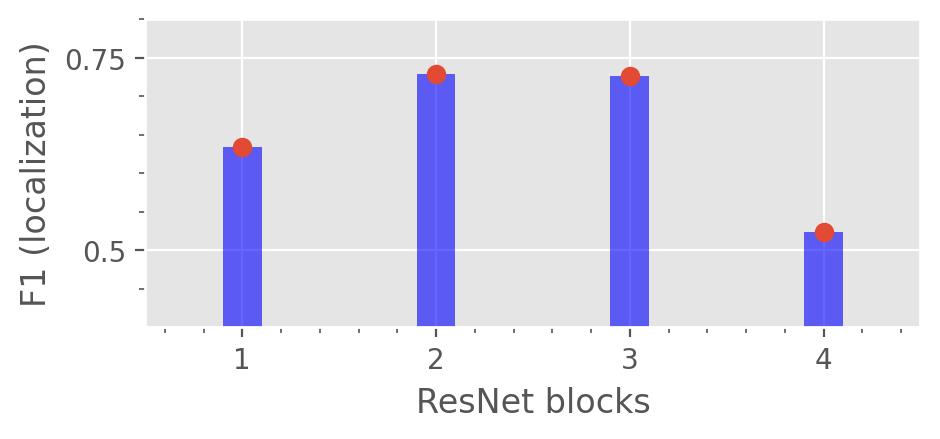}
    \end{center}
    \vspace{-2 em}
    \caption{Ablation study on the number of residual blocks used for feature extraction. Metric presented for mitosis localization.}
    \label{fig:blocks}
\end{figure}

\paragraph{\textbf{On the importance of the feature complexity}.} Convolutional neural networks combine stacked convolutional and pooling operations, which merge spatial information. Thus, later layers in CNNs extract high-level features with complex shapes, and low spatial resolution. Although CNNs for classification tasks usually benefit from deep structures, we observed that spatial resolution and low-level features are vital for mitosis localization, as shown in Figure \ref{fig:blocks}. For that reason, we used only 3 residual blocks of ResNet-18 architecture for the proposed method.

\vspace{-2 em}

\subsection{Qualitative evaluation}

\begin{figure}[htb]
    \begin{center}
    \includegraphics[width=0.8\linewidth]{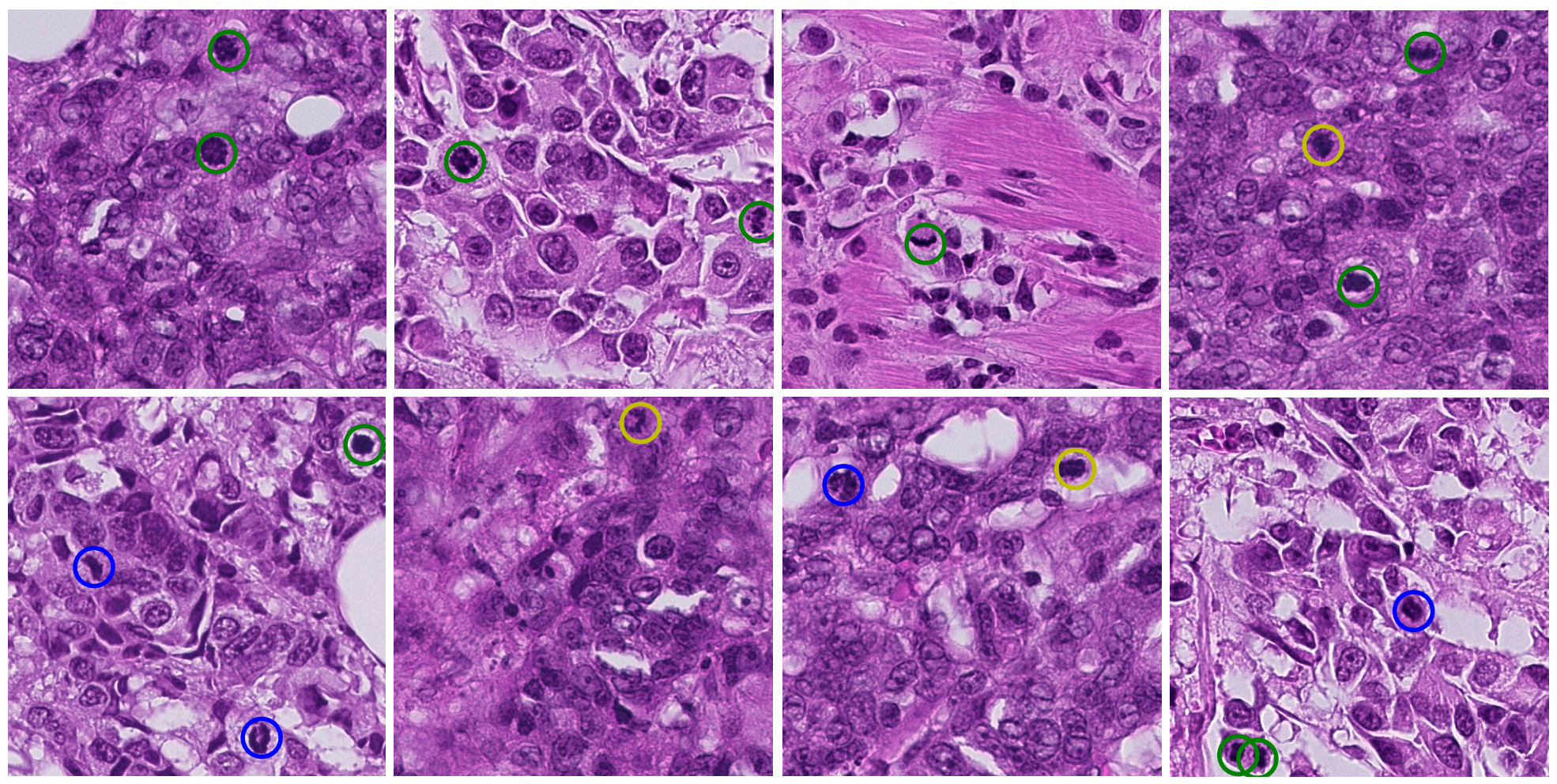}
    \end{center}
    \vspace{-2 em}
    \caption{Qualitative evaluation of the proposed method for mitosis localization. Green: true positive; Blue: false negative; Yellow: false positive.}
    \label{qualitative}
\end{figure}

Finally, we present visual results of the proposed method performance on the test subset in Figure \ref{qualitative}. In particular, correct detections of mitotic cells (true positives), cells wrongly classified as mitosis (false positives) and non-detected mitosis (false negatives) are shown in green, yellow and blue colors, respectively. Visual results show a promising performance of the proposed method, with false positive classifications occur with irregularly-shaped non-mitotic cells.

%Finally, we present visual results of the proposed method performance on the test subset in Figure \ref{qualitative}. In particular, we present results of true positive detections (\textit{top row}), cells wrongly classified as mitotic (\textit{bottom row, center}), and non-detected mitosis (\textit{bottom row, right}). Visual results show a promising performance of the proposed method, with false positive classifications occur with irregularly-shaped non-mitotic cells.

%% file: sections/6_conclusions.tex
\section{Conclusions}

In this work, we have presented a deep learning model for weakly supervised mitosis location on H\&E histology images. In particular, the model is composed of a narrow CNN backbone that leverages pixel-level predictions. Then, those predictions are grouped into an image-level score using maximum aggregation, that serves as proxy for CNN training via global labels. Thanks to the maximum operation, that only focus on very discriminative cells, obtained results have very few false positive predictions, and reaching a precision of $0.739$ and an F-score of $0.729$ on TUPAC16 dataset. The proposed approach, yet simple, reaches competitive performance in comparison to previous literature, without requiring any information of mitosis localization in the image during training. This calls into question the efficiency of other approaches, which require this location information, and resort to multiple phases of training to refine centroid-based labels and to alleviate false positive predictions. Further research could complement the proposed setting to take into account uncertainties on predicted mitoses, and to incorporate location information using a soft, constrained formulation.